\documentclass[journal]{IEEEtran}
\usepackage{cite}
\usepackage{bm}
\usepackage{multirow}
\usepackage[cmex10]{amsmath}
\usepackage{graphicx}
\usepackage{url}
\usepackage{color}
\usepackage{amsfonts}
\usepackage{amsmath}
\usepackage{extarrows}
\usepackage{graphicx}
\usepackage{amsfonts}
\usepackage{caption}
\usepackage{algorithm}
\usepackage{algpseudocode}
\usepackage{indentfirst}
\usepackage{caption}
\usepackage{esint} 
\usepackage{graphicx}
\usepackage{graphics}
\usepackage{balance}
\usepackage{epsfig}
\usepackage{soul}
\usepackage{caption}
\usepackage{gensymb}
 \DeclareMathOperator*{\argmax}{argmax}
\DeclareMathOperator*{\argmin}{argmin}

\makeatletter
\def\endthebibliography{%
  \def\@noitemerr{\@latex@warning{Empty `thebibliography' environment}}%
  \endlist
}
\makeatother

\def\bA{{\mathbf{A}}}  \def\bC{{\mathbf{C}}}  
 \def\bG{{\mathbf{G}}} \def\bH{{\mathbf{H}}}  
    
\def\bP{{\mathbf{P}}}    
  \def\bW{{\mathbf{W}}}  \def\bY{{\mathbf{Y}}}

 \def\bb{{\mathbf{b}}}   
\def\bf{{\mathbf{f}}}    
  \def\bm{{\mathbf{m}}} \def\bn{{\mathbf{n}}} 
  \def\br{{\mathbf{r}}}  \def\bt{{\mathbf{t}}}
  \def\bw{{\mathbf{w}}}  \def\by{{\mathbf{y}}}

\def\argmin{\mathop{\mathrm{argmin}}}

\begin{document}
\title{Adaptive Beamforming Design for mmWave RIS-Aided Joint Localization and Communication}

\author{Jiguang~He$^\dag$, Henk~Wymeersch$^\star$, Tachporn~Sanguanpuak$^\dag$, Olli Silv\'en$^\ddag$, Markku~Juntti$^\dag$\\
        $^\dag$Centre for Wireless Communications, FI-90014, University of Oulu, Finland\\
        $^\star$Department of Electrical Engineering, Chalmers University of Technology, Gothenburg, Sweden\\
        $^\ddag$Center for Machine Vision and Signal Analysis, FI-90014, University of Oulu, Finland\\
\thanks{This work has been performed in the framework of the IIoT Connectivity for Mechanical Systems (ICONICAL), funded by the Academy of Finland. This work is also partially supported by the Academy of Finland 6Genesis Flagship (grant 318927) and Swedish Research Council (grant no. 2018-03701). }}
 \maketitle
\begin{abstract}
The concept of reconfigurable intelligent surface (RIS) has been proposed to change the propagation of electromagnetic waves, e.g., reflection, diffraction, and refraction. To accomplish this goal, the phase values of the discrete RIS units need to be optimized. In this paper, we consider RIS-aided millimeter-wave (mmWave) multiple-input multiple-output (MIMO) systems for both accurate positioning and high data-rate transmission. We propose an adaptive phase shifter design based on hierarchical codebooks and feedback from the mobile station (MS). The benefit of the scheme lies in that the RIS does not require deployment of any active sensors and baseband processing units. During the update process of phase shifters, the combining vector at the MS is also sequentially refined. Simulation results show the performance improvement of the proposed algorithm over the random design scheme, in terms of both positioning accuracy and data rate. Moreover, the performance converges to exhaustive search scheme even in the low signal-to-noise ratio regime. 
\end{abstract}


\section{Introduction}
\label{sec:intro}
Recently, the concept of reconfigurable intelligent surface (RIS), also known as large intelligent surface (LIS), has been proposed to further improve energy efficiency (EE) and spectrum efficiency (SE). The RIS is capable of controlling the electromagnetic waves via meta-material units, passively in most of the cases, without any need of sophisticated baseband processing units and radio frequency (RF) chains~\cite{Huang2018, Ertugrul2019arXiv,Hu2018}. Unlike the conventional reflecting surfaces, the RIS can be regarded as a phased array, which is able to steer the beams of incident radio waves towards different directions and break the \textit{law of reflection}. Much more flexibility on reflection directions (possible any angles within $(-\frac{\pi}{2},\; \frac{\pi}{2})$) is offered by the RIS compared to its conventional counterpart.  

It was shown in~\cite{he2019large}, with the introduction of a RIS, significant enhancement on positioning error bound (PEB) and orientation error bound (OEB) can be realized even via a medium-sized RIS with the assumption of perfect channel station information (CSI). To optimally design the phase shifters at the RIS, the channels linking the base station (BS) and mobile station (MS) via the RIS need to be known \textit{a priori}. In~\cite{Taha2019}, the authors proposed two efficient approaches for the analog phase shifter design based on compressive sensing (CS) and deep learning (DL). However, a small portion of the RIS elements has to be made active. Meanwhile, computational complexity involved in the process cannot be neglected. 

It is reasonable that radio-based positioning resorted to millimeter wave (mmWave) multiple-input multiple-output (MIMO) systems~\cite{Shahmansoori2018, Abu-Shaban2018,Zhao2018, Koirala2018,Kakkavas2019} thanks to the high resolution on both angular and temporal domains. It was shown that even a single BS can achieve promising positioning accuracy. In~\cite{he2019large}, we studied the potentials of RIS for positioning by integrating it with mmWave MIMO systems and derived the fundamental performance limits based on Fisher information and Cram\'er-Rao lower bounds (CRLB). In such systems, the advantage is that the BS has prior information about the angle of departure (AoD), angle of arrival (AoA), and time of arrival (ToA) of the BS-RIS path upon the deployment of the RIS. This information can be directly utilized in the process of designing analog phase shifters at the RIS and beam alignment for the combiner at the MS . 

In this paper, we study the positioning and communication performance with the assistance of a RIS based reflector and multiple subcarriers at mmWave frequency bands. We focus on the design of analog phase shifters at the RIS based on hierarchical analog codebooks and limited feedback message from the MS to the RIS controller for indicating the optimal beam index. Also, the concept of hierarchical codebooks is applied to the design of combiner at the MS. We study the performance in terms of position error (PE), orientation error (OE) at the MS, and achievable data rate between the BS and MS. Numerical results show the superiority of the RIS-aided mmWave MIMO positioning and communication system with hierarchical codebooks over its counterpart with random phase shifters. Moreover, the performance of the proposed scheme converges to the exhaustive search approach even in the low signal-to-noise ratio (SNR) regime.  
\begin{figure}[t]
	\centering
	\includegraphics[width=0.9\linewidth]{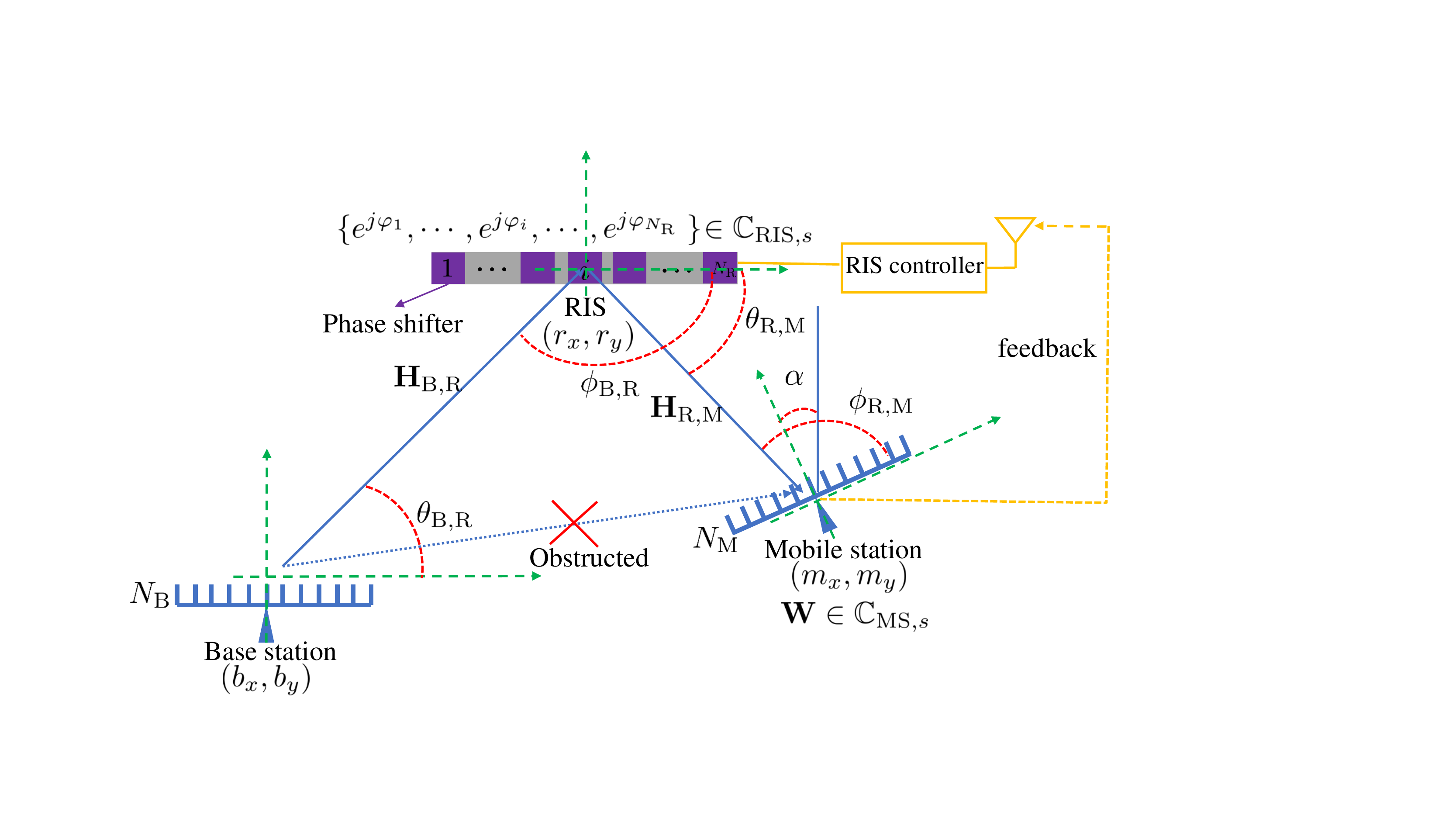}
	\caption{Positioning and communication system with the aid of a RIS and multi-carrier mmWave OFDM signals. The coordinates and orientation of the MS, $(m_x, m_y)$ and $\alpha$, are unknown and to be estimated with the LoS obstructed. Limited feedback is utilized in the design of analog phase shifters at the RIS and hybrid analog-digital combiners at the MS. Meanwhile, the system targets at high data-rate transmissions between the BS and the MS via the RIS. }
	\label{System_Model}
	\vspace{-0.5cm}
\end{figure}
\section{System Model}
The RIS-aided mmWave MIMO positioning and communication system is presented in Fig.~\ref{System_Model}, which consists of one multiple-antenna BS, one multiple-antenna MS and one RIS. We consider a two-dimensional (2D) scenario with uniform linear arrays (ULAs) for both the antenna elements and the RIS elements (i.e., discrete analog phase shifters). Extensions to uniform planar arrays (UPAs) and 3D positioning are feasible. The numbers of antenna elements at the BS and MS are $N_\text{B}$ and $N_\text{M}$, respectively, while the number of discrete elements at the RIS is $N_\text{R}$. Under the assumption of mmWave MIMO, we have the following constraint: $\{N_\text{B}, N_\text{M}\} \gg 1$. No rotation is assumed for the BS and the RIS while $\alpha$-rad rotation is assumed for the MS. The line-of-sight (LoS) link is susceptible to blockage. Therefore, we focus on the scenario where LoS is blocked. The objective of the system is to localize the MS, estimate its orientation, and enable high data-rate transmission by using the received signals at the MS with $N$ mmWave orthogonal frequency division multiplexing (OFDM) subcarriers when the LoS path is obstructed. 

The propagation channel connecting the BS to the MS is only composed of one reflection path via the RIS. The two tandem channels ($\bH_\text{B,R}[n] \in \mathbb{C}^{N_\text{R} \times N_\text{B}}$ for the first link and $\bH_\text{R,M}[n]\in \mathbb{C}^{N_\text{M} \times N_\text{R}}$ for the second link) for the $n$-th subcarrier, for $n = - (N-1)/2,\cdots, (N-1)/2$, defined as 
\begin{equation}
 \bH_\text{B,R}[n]=  \rho_\text{B,R} e^{-j2\pi\tau_\text{B,R} \frac{nB}{N}}\boldsymbol{\alpha}_{r}(\phi_\text{B,R})\boldsymbol{\alpha}_{t}^H(\theta_\text{B,R}), 
\end{equation} 
and  
\begin{equation}\label{H_LM}
\bH_\text{R,M}[n] =  \rho_\text{R,M} e^{-j2\pi\tau_\text{R,M} \frac{nB}{N}} \boldsymbol{\alpha}_{r}(\phi_\text{R,M}) \boldsymbol{\alpha}_{t}^H(\theta_\text{R,M}), 
\end{equation} 
where $\boldsymbol{\alpha}_{r}(\phi_\text{B,R}) \in \mathbb{C}^{N_\text{R} \times 1}$ and $\boldsymbol{\alpha}_{t}(\theta_\text{B,R}) \in \mathbb{C}^{N_\text{B} \times 1}$ are the (antenna) array response vectors at the RIS and BS, respectively. The $i$-th entry of $\boldsymbol{\alpha}_{r}(\phi_\text{B,R})$ and $\boldsymbol{\alpha}_{t}(\theta_\text{B,R})$ are $[\boldsymbol{\alpha}_{r}(\phi_\text{B,R})]_i = e^{j 2 \pi (i-1) \frac{d}{\lambda} \sin(\phi_\text{B,R})}$, $[\boldsymbol{\alpha}_{t}(\theta_\text{B,R})]_i = e^{j 2 \pi (i-1) \frac{d}{\lambda} \sin(\theta_\text{B,R})}$ with $d$ being the antenna element spacing\footnote{With notation reuse, $d$ also denotes element spacing in the RIS.}, $\lambda$ being the wavelength of the signal, and $\phi_\text{B,R}$ and $\theta_\text{B,R}$ being the AoA and AoD for the BS-RIS channel, respectively. $\boldsymbol{\alpha}_{t}(\theta_\text{R,M})$ and $\boldsymbol{\alpha}_{r}(\phi_\text{R,M})$ in~\eqref{H_LM} are defined in the same manner as $\boldsymbol{\alpha}_{t}(\theta_\text{B,R})$ and $\boldsymbol{\alpha}_{r}(\phi_\text{B,R})$. $j = \sqrt{-1}$, $\tau_\text{B,R}$ is the ToA for the BS-RIS link, $B$ is the overall bandwidth for all the subcarriers, and $B \ll f_c$\footnote{All the wavelengths $\lambda$'s of the subcarriers are nearly the same because of $B \ll f_c$ in mmWave communications.}, where $f_c$ is the center frequency. $\rho_\text{B,R} \in \mathbb{R}^+$ is related to the free-space path loss occurred in the first link for all the subcarriers, and $(\cdot)^H$ denotes the conjugate transpose operation. $\rho_\text{R,M}$ and $\tau_\text{R,M}$ in~\eqref{H_LM} are defined in the same way as $\rho_\text{B,R}$ and $\tau_\text{B,R}$.

The entire channel between the BS and the MS via the RIS for the $n$-th subcarrier can be formulated as
\begin{equation}\label{RIS_Channel1}
\bH[n] =  \bH_\text{R,M}[n] \boldsymbol{\Phi} \bH_\text{B,R}[n],
\end{equation}  
where $\boldsymbol{\Phi} =\frac{1}{\sqrt{N_\text{R}}} \mathrm{diag}(e^{j \varphi_1} ,\cdots, e^{j \varphi_{N_\text{R}}} )  \in \mathbb{C}^{N_\text{R} \times N_\text{R}}$ is the phase control matrix at the RIS. It is a diagonal matrix with constant-modulus entries in the diagonal and assumed to be independent of subcarrier index $n$. Setting $\beta = \boldsymbol{\alpha}^H_{t}(\theta_\text{R,M}) \boldsymbol{\Phi}\boldsymbol{\alpha}_{r}(\phi_\text{B,R}) = \left[\boldsymbol{\alpha}_{t}(\theta_\text{R,M}) \odot \boldsymbol{\alpha}^*_{r}(\phi_\text{B,R})\right]^H \boldsymbol{\varphi}$ with  $\boldsymbol{\Phi}=\mathrm{diag}(\boldsymbol{\varphi})$ and $\odot$ being element-wise product. The channel in~\eqref{RIS_Channel1} can be further expressed as 
\begin{equation}\label{RIS_Channel2}
\bH[n] = \beta \rho_\text{B,R}  \rho_\text{R,M} e^{-j2\pi (\tau_\text{B,R} + \tau_\text{R,M}) \frac{nB}{N}} \boldsymbol{\alpha}_{r}(\phi_\text{R,M}) \boldsymbol{\alpha}_{t}^H(\theta_\text{B,R}). 
\end{equation}  
It is certain that $\bH[n]$ is a rank-1 matrix based on~\eqref{RIS_Channel2}. Intuitively, in the proposed system, the beam at the BS should be steered towards the RIS. Meanwhile, the beam at the RIS should be designed towards the MS. It is reasonable to assume that the center of RIS is known to the BS. Therefore, the beam can be made very sharp based on the \textit{a priori} AoD information (i.e., $\theta_\text{B,R}$, calculated based on the known centers of the BS and the RIS) at the BS. However, the design of $\boldsymbol{\Phi}$ is not straightforward. The optimal $\boldsymbol{\varphi}^*$ is obtained as 
\begin{equation}
\boldsymbol{\varphi}^* = \argmax_{\boldsymbol{\varphi}} |\beta| = \frac{1}{\sqrt{N_\text{R}}} \boldsymbol{\alpha}_{t}(\theta_\text{R,M}) \odot \boldsymbol{\alpha}^*_{r}(\phi_\text{B,R}), 
\end{equation}
with $|\beta| = \sqrt{N_\text{R}}$, provided that $\theta_\text{R,M}$ and $\phi_\text{B,R}$ are known. In the proposed system, we know $\phi_\text{B,R}$, which is equal to $-\pi + \theta_\text{B,R}$. However, the information on $\theta_\text{R,M}$ is unknown and supposed to be estimated. The other unknown parameters in~\eqref{RIS_Channel2} are $\phi_\text{R,M}$ and $\tau_\text{R,M}$. The estimates of these channel parameters can be used to locate the MS' coordinate and orientation. Also, they can be used to enable high data-rate transmission by guiding the design of optimal or suboptimal combiner at the MS. 

We fix the beamforming vector at the BS as $\mathbf {f} = \frac{1}{\sqrt{N_\text{B}}} \boldsymbol{\alpha}_{t}(\theta_\text{B,R})$ and $\|\mathbf {f}\|_2 = 1$, where $\|\cdot\|_2$ stands for the Euclidean norm. The positioning reference signal (PRS) $x[n]$ with $|x[n]| =1$ is transmitted over the $n$-th subcarrier, the downlink received signal is in the form of 
\begin{equation}\label{Received_signal}
\by[n] = \sqrt{P}\bH[n] \mathbf{f} x[n] +  \bn[n],
\end{equation} 
where each entry in the additive white noise $\bn[n]$ follows circularly-symmetric complex normal distribution $\mathcal{CN}(0, \sigma^2)$, and $P$ is the transmit power of the PRS. 
After post-processing, the received signal in~\eqref{Received_signal} can be further written as 
\begin{equation}\label{Received_signal_postprocessing}
\bar{\by}[n] = \bW^H\by[n] =\bW^H (\sqrt{P}\bH[n]  \mathbf{f} x[n] +  \bn[n]),
\end{equation}
where $\bW = \bW_{\text{RF}} \bW_{\text{BB}}$ is the hybrid combiner with $\bW_{\text{RF}}$ and $ \bW_{\text{BB}}$ being the analog and digital combiners, respectively. 

\section{Hierarchical Codebook Design}
\label{Codebook_design}
Different types of hierarchical codebooks are leveraged at the RIS and the MS due to different architectures. The RIS consists of purely-analog phase shifters, thus, its hierarchical codebooks only contain analog codewords accordingly. The MS is assumed to have a hybrid analog-digital architecture. Therefore, the hierarchical codebooks at the MS contain codewords in a hybrid format. Nevertheless, the design of hierarchical codebooks can be summarized in the following three major steps:
\begin{itemize}
\item Solve the equation with least squares (LS) $ \bA^H \bC_s= \bG_s$~\cite{Alkhateeb2014}, where $\bA$, $\bC_s$, and $\bG_s$ are defined below.

\item Normalize each column in $\bC_s$, so that $\| [\bC_s]_{:,m}\|_2 =1$, $\forall m$, where $[\bC_s]_{:,m}$ denotes the $m$-th column of $\bC_s$.

\item Make each column in $\bC_s$ satisfy the hardware constraint~\cite{Tranter2017,Yu2016}. 
\end{itemize} 

\begin{figure}[t]
	\centering
	\includegraphics[width=0.6\linewidth]{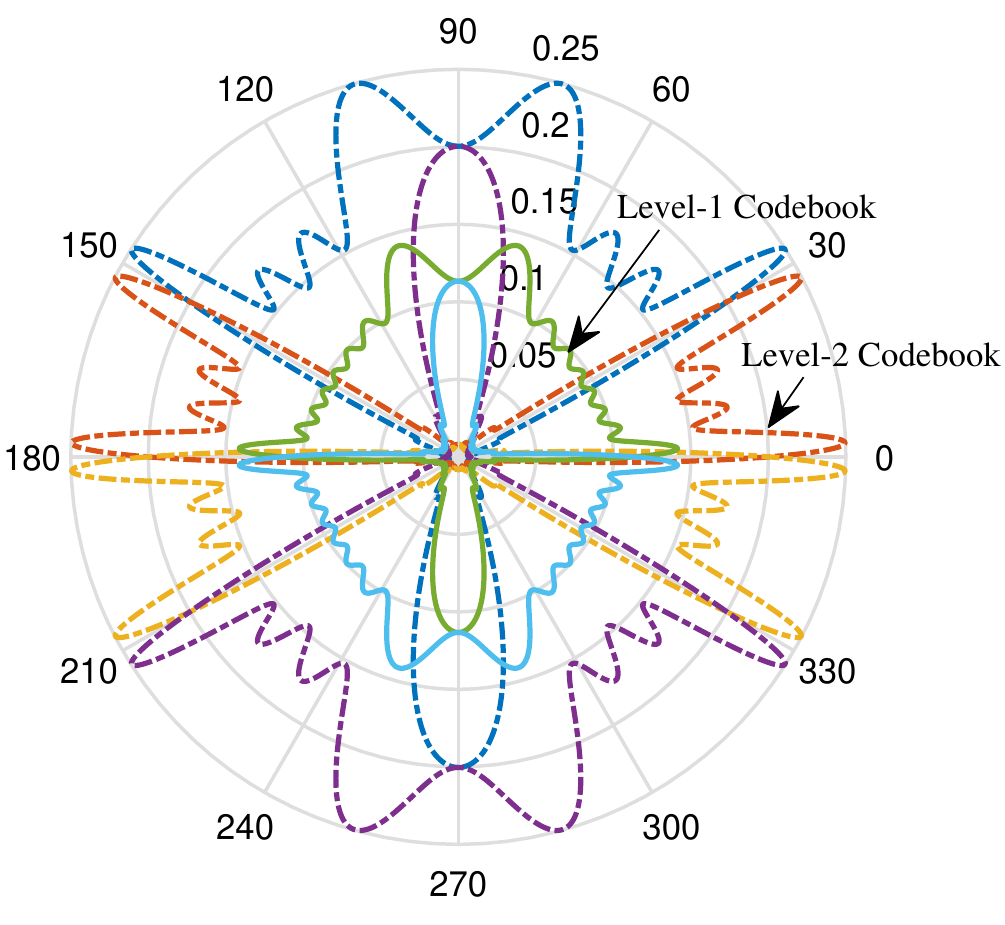}
	\caption{An example on hierarchical codebook consisting of 2 levels. The level-1 codebook contains 2 codewords (plotted in solid line), and the level-2 codebook contains 4 codewords (plotted in dotted line). }
	\label{Hierarchical_codebook}
\end{figure}

In the first step, the $(n,m)$-entry of $\bA$ is $[\bA]_{n,m} = e^{ j 2\pi \frac{d}{\lambda} (n -1) \bar{\theta}_m}$, for $m = 1, \cdots, M$ and $n = 1,\cdots, N_\text{B}(N_\text{R})$, with $\bar{\theta}_m = -1 + \frac{2m - 1}{M}$, where $M$ is the total number of quantized levels for interval $[-1, 1]$.\footnote{We assume that $M \geq \{N_\text{B}, N_\text{R}\}$. A larger $M$ results in smoother beam pattern of the codewords, obtained in the first step. Also, instead of quantizing the phase $[-\pi, \pi]$, we quantize the range of the sinusoidal function $[-1, 1]$.} Each column of $\bC_s$ is a individual codeword in the level-$s$ codebook, where $s = 1, \cdots, S$. The number of codewords increases exponentially as a function of level index $s$, i.e., $K^s$, where $K$ is the number of codewords at the first level. $\bG_s$ is a matrix with binary elements and obtained by circular shift operation. For instance, in the first level, the first column of $\bG_s$ is $[\bG_s]_{:,1} = [\mathbf{1}^T \; \mathbf{0}^T ]^T$, where the number of 1's is $M/K$ (assumed to be a positive integer without loss of generality). The remaining $K-1$ columns are obtained by shifting $M/K$ elements down sequentially. Constructing $\bG_s$ in such a way is to make the beam pattern of $\bC_s$ constant over a certain range of angle. The range of angle reduces as the level of codebook increases. In other words, the resolution of codewords sequentially increases. An example is provided in Fig.~\ref{Hierarchical_codebook} to illustrate the relationship between the codewords in different levels. 
After obtaining $\bC_s = (\bA^H)^\dagger \bG_s$ by LS with $(\cdot)^\dagger$ denoting Moore-Penrose pseudo inverse, the next steps are to normalize each codeword to $1$ and make it customized for the architecture. Different design methods are considered at the RIS and the MS, detailed in the sequel.  

\subsection{RIS}
For the RIS, each element in the codewords has to be constant modulus, i.e., $|[\bar{\bC}_{\text{RIS},s}]_{m,n}| = \frac{1}{\sqrt{N_\text{R}}}$, $\forall m, n$. Here, we add ``RIS'' to the subscript to make it specific. In order to find the corresponding codebooks that satisfy the hardware constraint, the optimization problem is formulated as 
\begin{align}\label{RIS_step3}
[\bar{\bC}_{\text{RIS},s}]_{:,n} = \argmin\|[\hat{\bC}_{\text{RIS},s}]_{:,n} -[\bC_{\text{RIS},s}]_{:,n} \|_2^2, \nonumber\\
\text{s.t.}\;\; |[\hat{\bC}_{\text{RIS},s}]_{m,n}| =  1/\sqrt{N_\text{R}},
\end{align}
which is carried out in the column-by-column manner. The potential algorithm to address this problem is gradient projection~\cite[Algorithm 1]{Tranter2017}. It mainly comprises two steps: 1) gradient descent, 2) projection onto a constant modulus space. More details can be found in~\cite{Tranter2017}.    

\subsection{MS}
For the MS, its hardware architecture is different from that of RIS. Assuming that there exist $N_{\text{RF}}$ RF chains at the MS, maximal $N_{\text{RF}}$ observations can be accessed at the MS and $N_{\text{RF}}$ codewords can be used as the combiner simultaneously. To make $\bC_s$ obtained in the second step suitable for the MS, we need to address the matrix factorization problem,  
\begin{align}\label{matrix_factorization}
\{\bar{\bC}_{ \text{MS, RF}, s,k}, \bar{\bC}_{\text{MS, BB}, s, k}   \} &= \argmin\| \hat{\bC}_{\text{MS, RF}, s,k}\hat{\bC}_{ \text{MS, BB}, s, k}\nonumber\\ &-[\bC_s]_{:, (k-1) N_{\text{RF}} +1 : k N_{\text{RF}} }\|_F^2, \nonumber\\
&\text{s.t.}\;[\hat{\bC}_{ \text{MS, RF}, s, k}]_{m,n} =  \frac{1}{\sqrt{N_\text{M}}},\nonumber\\
&\|\hat{\bC}_{\text{MS, RF}, s,k}\hat{\bC}_{ \text{MS, BB}, s,k} \|_F^2 = N_{\text{RF}},
\end{align}
where $[\bar{\bC}_ {\text{MS},s}]_{:, (k-1) N_{\text{RF}} +1 : k N_{\text{RF}} } = \bar{\bC}_{ \text{MS, RF}, s,k} \bar{\bC}_{\text{MS, BB}, s, k}$, for $k = 1, \cdots, K^s/N_{\text{RF}}$ and $\|\cdot\|_F$ denotes the Frobenius norm. Unlike the optimization problem in~\eqref{RIS_step3} for the RIS, the optimization problem for the MS is carried out for $N_{\text{RF}}$ columns per time. The matrix factorization shown in~\eqref{matrix_factorization} can be solved by alternating minimization using phase extraction (PE-AltMin)~\cite{Yu2016}. The core idea is to fix one variable and optimize the other iteratively until a certain stopping criterion is met. 

Each column of $\bar{\bC}_{\text{RIS},s}$ and $\bar{\bC}_{\text{MS},s}$ obtained after going through all the three steps, for $s = 1, \cdots, S$, is regarded as an individual codeword for the level-$s$ codebook of the RIS and MS, respectively.  

\section{Adaptive Phase Shifter Design}
\subsection{Proposed Protocol}
\begin{figure}[t]
	\centering
	\includegraphics[width=0.9\linewidth]{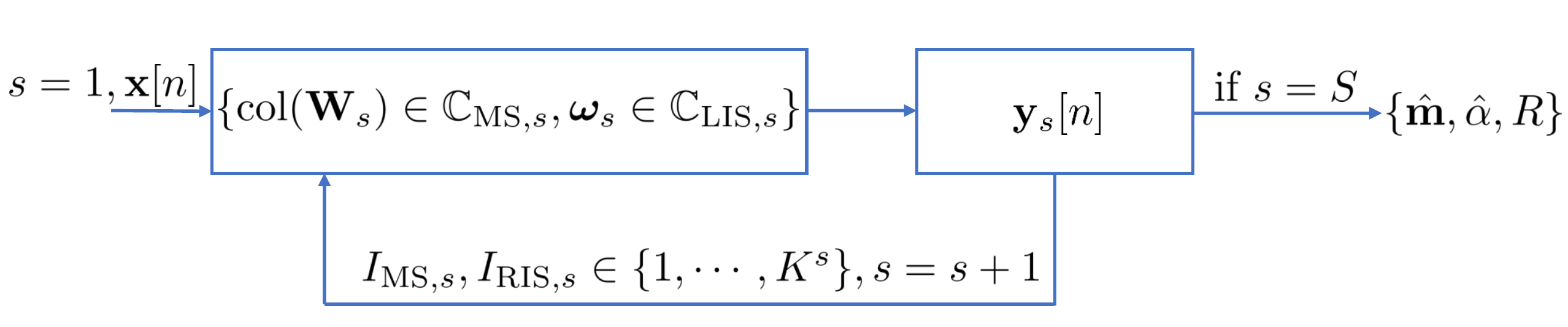}
	\caption{Adaptive update scheme for the phase shifters design based on the feedback and receive beam alignment for the purpose of both positioning and communications.}
	\label{Update_Process}
\end{figure}

The combining beam vectors/matrix $\bW$ in~\eqref{Received_signal_postprocessing} and $\boldsymbol{\varphi}$ are selected from the multi-resolution hierarchical codebooks in an adaptive manner. For the purpose of better illustration, we differentiate the codebooks of the MS and RIS by defining the following ``sets of sets'': $\mathbb{C}_{\text{MS}} = \{\mathbb{C}_{\text{MS}, 1}, \cdots, \mathbb{C}_{\text{MS}, S}\}$ and $\mathbb{C}_{\text{RIS}} = \{\mathbb{C}_{\text{RIS}, 1}, \cdots, \mathbb{C}_{\text{RIS}, S}\}$, which are obtained according Section~\ref{Codebook_design}. The columns of $\bar{\bC}_{\text{RIS},s}$ forms $\mathbb{C}_{\text{RIS}, s}$, and the same for $\bar{\bC}_{\text{MS},s}$. The cardinality of $\mathbb{C}_{\text{MS},s}$ and $\mathbb{C}_{\text{RIS},s}$ are $|\mathbb{C}_{\text{MS},s}| = |\mathbb{C}_{\text{RIS},s}| = K^s$, for $s = 1, \cdots, S$. The received signal at stage $s$ is defined as  
\begin{align}
\by_s[n] &=  \bW_s^H (\sqrt{P}\bH[n]  \mathbf{f} x[n] +  \bn[n]), \nonumber\\
 &\overset{(a)}{=} \sqrt{P N_\text{B}} \rho_\text{B,R}  \rho_\text{R,M} e^{-j2\pi (\tau_\text{B,R} + \tau_\text{R,M})\frac{nB}{N}} \beta (\boldsymbol{\varphi}_s) \bW_s^H \boldsymbol{\alpha}_{r}(\phi_\text{R,M}) \nonumber\\
 & +  \bW_s^H\bn[n], \label{Received_signal_training}
\end{align}
where $\mathrm{col}(\bW_s) \in \mathbb{C}_{\text{MS},s}$ and $\boldsymbol{\varphi}_s \in \mathbb{C}_{\text{RIS},s}$ with $\mathrm{col}(\bA)$ denoting the columns of matrix $\bA$. $(a)$ is obtained by setting $x[n] =1$ and $\bf =  \frac{1}{\sqrt{N_\text{B}}} \boldsymbol{\alpha}_{t}(\theta_\text{B,R})$.  In~\eqref{Received_signal_training}, we write $\beta$ as a function of $\boldsymbol{\varphi}_s$. A framework is provided in Fig.~\ref{Update_Process} to depict the adaptive scheme in the proposed RIS-aided mmWave MIMO systems. At stage $s$, only $K$ codewords from $\mathbb{C}_{\text{MS},s}$ and $K$ codewords from $\mathbb{C}_{\text{RIS},s}$ are considered. 

Let us define the received signal matrix $\bY_s[n] \in \mathbb{C}^{K\times K}$ and associated sum-power matrix $[\bP_s]_{m,k} = \sum\limits_{n =1}^N |[\bY_{s}[n]]_{m,k}|^2 $, where $[\bY_{s}[n]]_{m,k}$ corresponds to the received signal associated with the $m$-th selected codeword from $\mathbb{C}_{\text{MS},s}$ and the $k$-th selected codeword from $\mathbb{C}_{\text{RIS},s}$. Then, we find the column and row indices of $\bP_s$ which corresponds to the largest value among the $K^2$ entries. The MS feeds back the index of codeword from $\mathbb{C}_{\text{RIS},s}$ (i.e., column index of $\bP_s$, defined as $I_{\text{RIS},s}$ in Fig.~\ref{Update_Process}) to the RIS controller. The RIS selects the $K$ associated codewords from $\mathbb{C}_{\text{RIS},s+1}$ as candidate analog beamformers in the next stage. Meanwhile, the MS picks up $K$ codewords from $\mathbb{C}_{\text{MS},s+1}$ for the next stage according to the row index of the codeword with the highest sum power (defined as $I_{\text{MS},s}$ in Fig.~\ref{Update_Process}). After reaching the final stage, estimation on the MS coordinate $\bm = [m_x \; m_y]^T$ and MS's orientation $\alpha$, and computation of data rate $R$ is conducted.

The number of time slots for each stage is $\frac{K^2}{N_\text{RF}}$ with total $\frac{S K^2}{N_\text{RF}}$ for the entire process. The denominator means that $N_\text{RF}$ combining beams can be used simultaneously during the process of phase shifter design. The overall number of consumed time slots depends on the number of levels of the hierarchical codebooks, the number of RF chains at the MS, and the size of the first codebook. For instance, if there are 2 RF chains available at the MS and the hierarchical codebooks have 6 levels with the cardinality of the initial codebook being 2, the overall number of time slots is 12. This is in general quite small compared to the number of antennas at the BS and MS. 

\subsection{Positioning and Communication}
For the positioning, we follow the two-stage positioning approach~\cite{Shahmansoori2018}. We estimate the channel parameters in the first stage, and then estimate the location of the MS accordingly. The selected codewords from $\mathbb{C}_{\text{RIS},S}$ and $\mathbb{C}_{\text{MS},S}$ and their associated received signals across all the subcarriers contain information about the channel parameters, e.g., $\theta_{\text{R,M}}$,  $\phi_{\text{R,M}}$, $\tau_{\text{R,M}}$. Their estimates are written as follows:
\begin{equation}
\hat{\theta}_\text{R,M} = \argmax_{\theta_\text{R,M}} | \left[\boldsymbol{\alpha}_{t}(\theta_\text{R,M}) \odot \boldsymbol{\alpha}^*_{r}(\phi_\text{B,R})\right]^H \boldsymbol{\varphi}_{\text{opt}}|, \label{est_theta}
\end{equation}
\begin{align}
\hat{\phi}_\text{R,M} &= \argmax_{\phi_\text{R,M} } | \bw_{\text{opt}}^H  \boldsymbol{\alpha}_t(\phi_\text{R,M} ) |, \label{est_phi} \\
\hat{\tau}_{\text{R,M}} &= \argmax_{\tau_{\text{R,M}}} |\by_S^H  \bt(\tau_{\text{R,M}}) | - \tau_\text{B,R}, \label{est_tau}
\end{align}
where  $\bt(\tau_{\text{R,M}})  = [e^{-j2\pi \frac{B\tau_{\text{R,M}}}{N} }, \cdots,  e^{-j2\pi B \tau_{\text{R,M}}}]^T$, $\boldsymbol{\varphi}_{\text{opt}}$ and $\bw_{\text{opt}}$ are selected from $\mathbb{C}_{\text{RIS},S}$ and $\mathbb{C}_{\text{MS},S}$ based on sum-power matrix $\bP_S$. $\by_S$ is formulated by stacking the $N$ received signal from all the subcarriers into a vector, which corresponds to the highest sum power at the last stage. Note that $\phi_\text{B,R}$ in~\eqref{est_theta}, $\tau_\text{B,R}$ in~\eqref{est_tau} are known, calculated as 
\begin{align}
&\tau_\text{B,R} =  \|\bb - \br \|_2/c,\nonumber\\
&\phi_\text{B,R} = -\pi+\arccos((r_x - b_x)/ \|\bb -\br\|_2),\nonumber
\end{align} 
where $\bb = [b_x \; b_y]^T$ and $\br = [r_x \; r_y]^T$ are the known centers of the BS and the RIS, respectively. 

Based on the estimates from~\eqref{est_theta}--\eqref{est_tau}, we can estimate the positioning information of the MS by using the geometry, formulated as 
\begin{align}\label{Geometry_relationship}
&\hat{\bm} = \br + c \hat{\tau}_{\text{R,M}} [\cos(\hat{\theta}_\text{R,M}) \; \sin(\hat{\theta}_\text{R,M}) ]^T, \\
&\hat{\alpha} =  \pi +\hat{\theta}_\text{R,M} - \hat{\phi}_\text{R,M},
\end{align}
where $c$ is the speed of light. It should be noted that we consider far-field communications. Therefore, additional constraints are imposed to the number of RIS elements: $\frac{2 (N_\text{R} d)^2}{\lambda} < \| \bb -\br\|_2$ and $\frac{2 (N_\text{R} d)^2}{\lambda} < \| \br -\bm\|_2$, which can be summarized as $N_\text{R} < \frac{\sqrt{\lambda}}{\sqrt{2} d} \cdot \min\{ \sqrt{\| \bb -\br\|_2}, \sqrt{ \| \br -\bm\|_2}\}$.

The performance metrics used here are positioning mean square error (MSE), orientation MSE, and achievable data reate, defined as 
\begin{align}
P_e & = E\{\|\bm - \hat{\bm}\|^2\}, \\
O_e & = E\{|\alpha-\hat{\alpha}|^2\}, \\
R &=  \sum\limits_{n =1}^N\log_2 (1 + \frac{P}{\sigma^2} |\bw_\text{opt}^H \bH[n]|_{\boldsymbol{\varphi} = \boldsymbol{\varphi}_{\text{opt}}} \mathbf{f}|^2), \label{Rate}
\end{align}
where $\bH[n]|_{\boldsymbol{\varphi} = \boldsymbol{\varphi}_{\text{opt}}}$ means $\boldsymbol{\varphi}$ in $\bH[n]$ equals $\boldsymbol{\varphi}_{\text{opt}}$.

\subsection{Asymptotic Performance Analysis}
As the received SNR increases, the estimation performance will saturate due to the finite-resolution of the level-$S$ codebook at the RIS and MS. It highly probable that we always find the right codewords for analog phase shifters at the RIS and combiner at the MS. The problem can be interpreted in the following way. First, we find the codewords in the last codebooks (i.e., $\mathbb{C}_{\text{RIS},S}$ and $\mathbb{C}_{\text{MS},S}$) which result in the highest sum power. Second, we obtain the estimates of channel parameters by following~\eqref{est_theta}--\eqref{est_tau}, where $\by_S$ is replaced by its noise-free counterpart. Last, we calculate the mean square errors of channel parameters, location, and orientation, which are regarded as theoretical lower bounds by using certain specific hierarchical codebooks.  

\section{Simulation Results}
The parameters are set up as follows: $(b_x, b_y )= (0,0)$, $(r_x, r_y) = (40, 60)$, $(m_x, m_y) =(60, 45)$, all in meters, $\alpha = \pi/10$, $\mu = 2.08$ (path loss exponent), $\rho_\text{B,R} =  (\|\bb -\br\|_2)^{-\mu/2}$, $\rho_\text{R,M} =  (\|\br -\bm\|_2)^{-\mu/2}$, $N_\text{B} =64$, $N_\text{M} = 16$,  $N_\text{RF} = 2$, $N = 31$, $B = 100$ MHz, $f_c = 60$ GHz, and $\alpha = \pi/10$. According to the far field constraints, $N_\text{R} \leq 100$. In the following experiments, we set $N_\text{R}=16$. The SNR is defined as $\frac{P \rho^2_{\text{B,R}}\rho^2_{\text{R,M}}}{\sigma^2}$, where $\sigma^2 = -174\; \text{dBm} + 10 \log_{10}(B/N) = -109$ dBm. For the purpose of comparison, we introduce two benchmarks: 1) random phase at RIS and level-$S$ codebook at MS, 2) exhaustive search using the level-$S$ codebooks at both the MS and the RIS. The parameters for the hierarchical codebooks are set as: $S = 6$, $K =2$. 
\subsection{Channel Parameter Estimation}
\begin{figure}[t]
	\centering
	\includegraphics[width=0.8\linewidth]{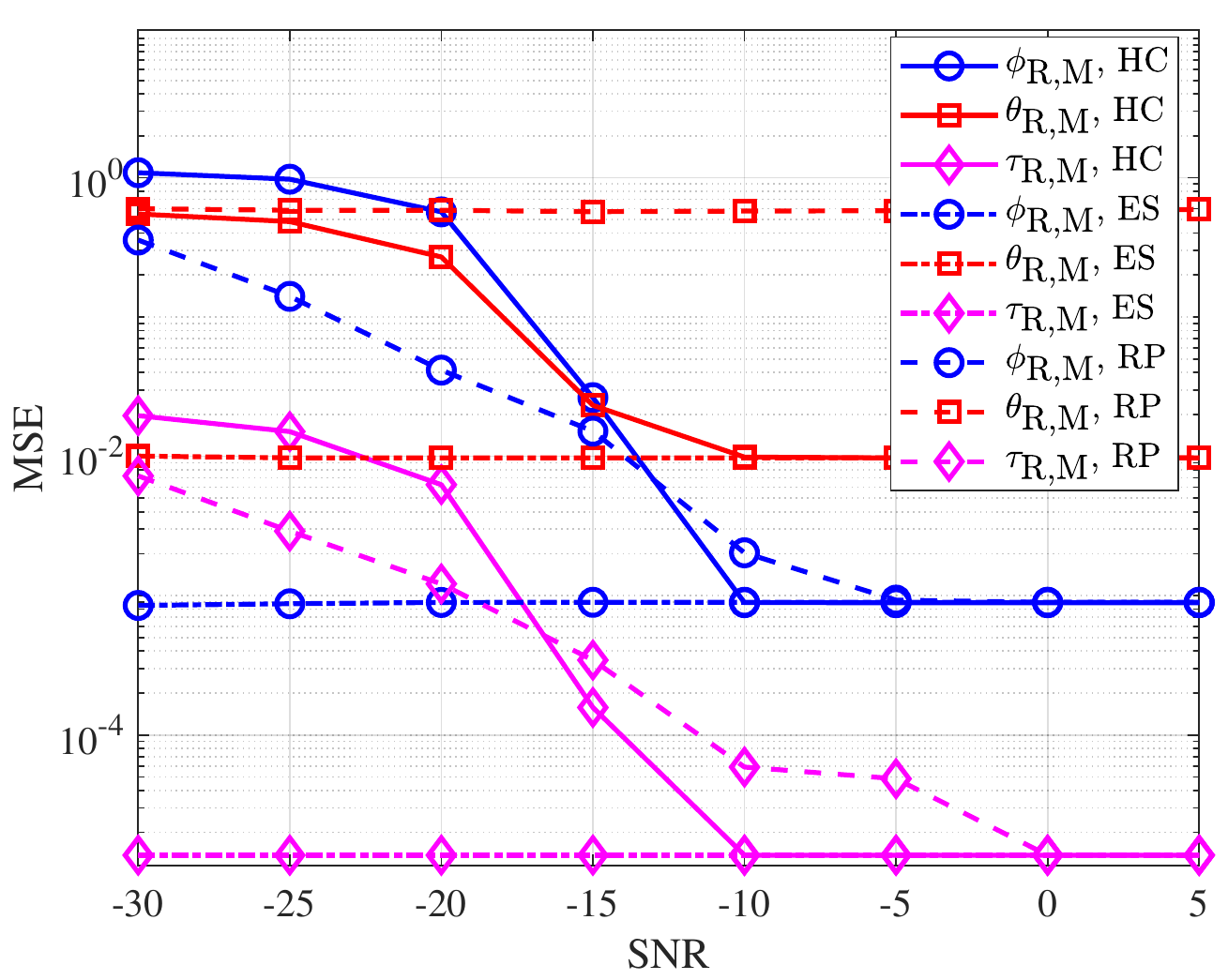}
	\caption{Channel parameter estimation for hierarchical codebooks, exhaustive search, and random phase.}
	\label{Channel_parameter_estimation}
	\vspace{-0.5cm}
\end{figure}

In this subsection, we focus on the evaluation of channel parameter estimation based on \eqref{est_theta}--\eqref{est_tau}. Exhaustive scheme algorithm based on the level-$S$ codebooks is considered as a benchmark scheme but with more time slots consumption (in total $64 \times 64/ 2 = 2048$). However, the number of required time slots for the proposed scheme is only $12 \ll 2048$. Another benchmark scheme with random phase and level-$S$ codebook consumes $K^S/N_\text{RF} = 32$ time slots. The simulation results are provided in Fig.~\ref{Channel_parameter_estimation}, where ``HC'', ``ES'', and ``RP'' in the legend denote hierarchical codebooks, exhaustive search, and random phase, respectively. The exhaustive search approach outperforms the one using hierarchical codebooks with significant increase on the training overhead. However, the performance of the proposed scheme converges to that of exhaustive search even in the low SNR regime. The theoretical lower bound is aligned with the exhaustive search. Therefore, we omit it in the simulation figure. The estimation performance of $\theta_\text{R,M}$ is worse than that of $\phi_\text{R,M}$ due to the difference of the constructed hierarchical codebooks. The beam pattern of $\mathbb{C}_{\text{MS}}$ is better than that of $\mathbb{C}_{\text{RIS}}$ thanks to the introduction of multiple RF chains at the MS. For the estimate of $\theta_\text{R,M}$, the random phase scheme is the worst among the three. Because of its randomness, no information about $\theta_\text{R,M}$ can be inferred from it.  

\subsection{PE and OE}
\begin{figure}[t]
	\centering
	\includegraphics[width=0.8\linewidth]{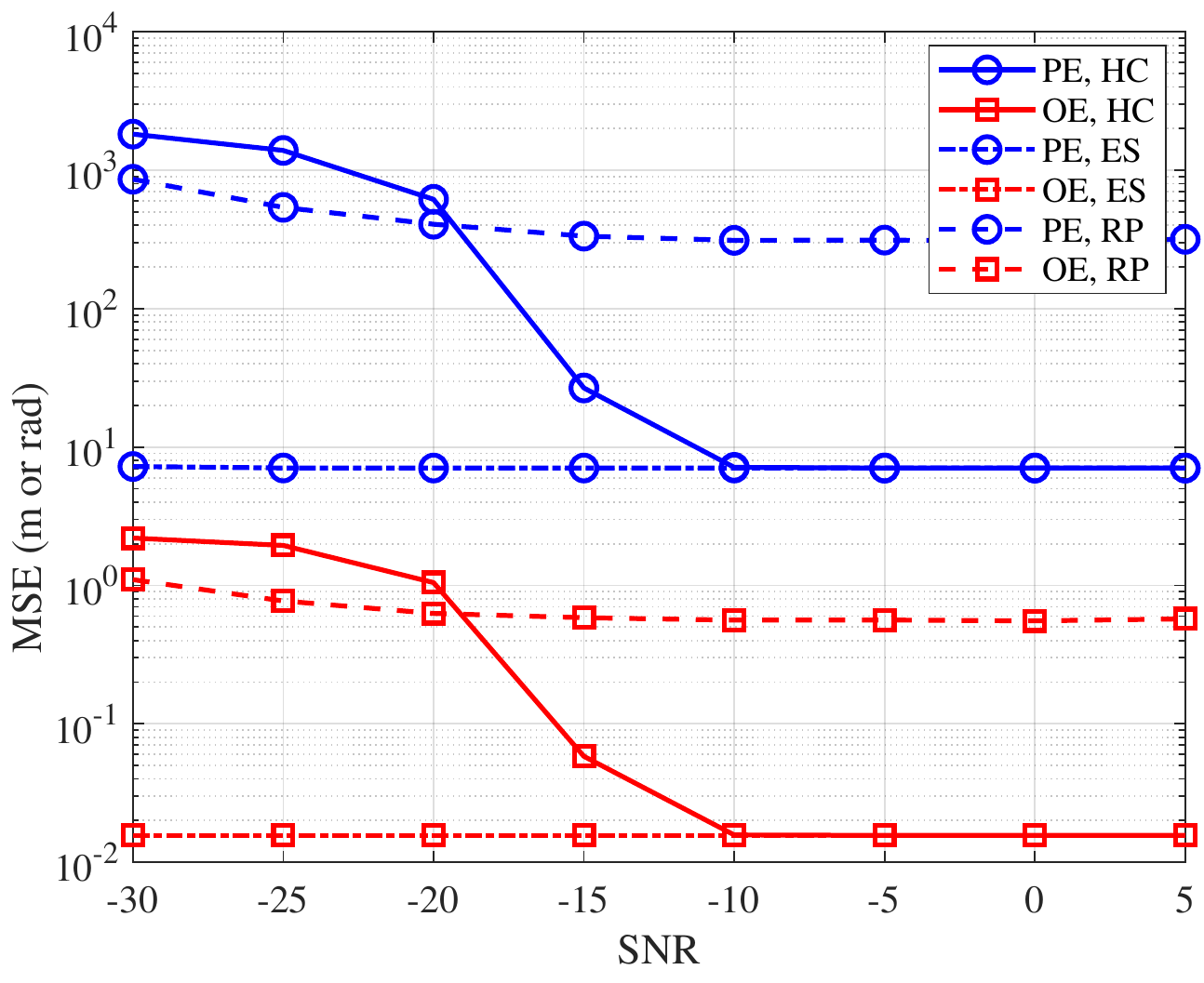}
	\caption{PE and OE for hierarchical codebooks, exhaustive search, and random phase.}
	\label{PEB_OEB}
	\vspace{-0.5cm}
\end{figure}

In this subsection, we provide the simulation results for both positioning error (PE) and orientation error (OE) by following the same parameter setup in the last subsection. We estimate the location of the MS based on the estimated channel parameters, i.e., $\hat{\theta}_\text{R,M}$ and $\hat{\tau}_\text{R,M}$. The simulation results are shown in Fig.~\ref{PEB_OEB}, which are consistent with the results in Fig.~\ref{Channel_parameter_estimation}. The performance of exhaustive search outperforms the proposed scheme with significant overhead. Again, the performance of the random phase is the worst among the three, which indicates that the phase of the RIS units plays a critical role in the positioning. 

\subsection{Achievable Rate}
\begin{figure}[t]
	\centering
	\includegraphics[width=0.8\linewidth]{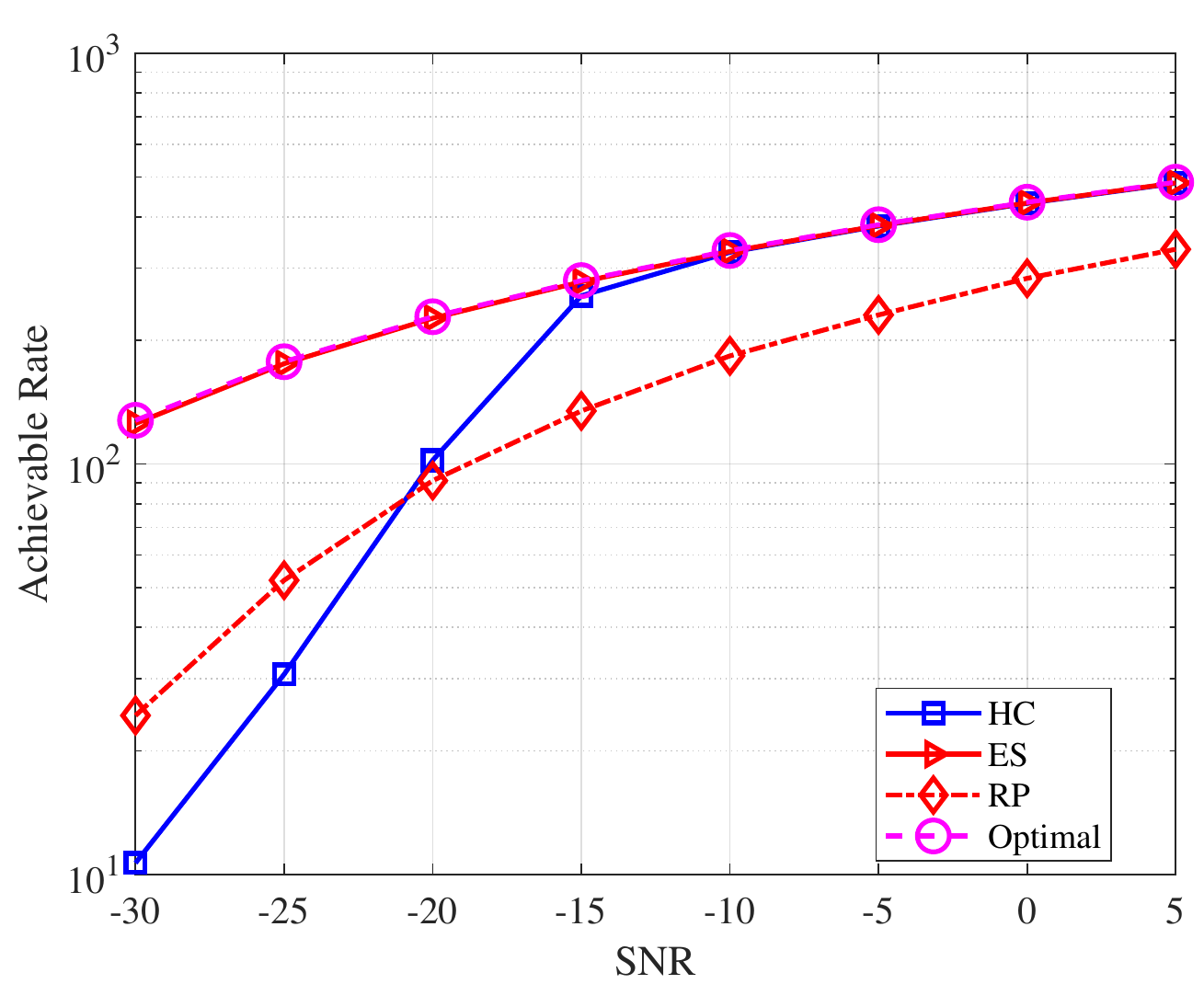}
	\caption{Achievable rate for hierarchical codebooks, exhaustive search, and random phase.}
	\label{Data_Rate}
\end{figure}

In this subsection, we consider the achievable rate for the RIS-aided mmWave MIMO systems according to~\eqref{Rate}. The performance is also comparable to the exhaustive search in the SNR regime $[-10 , \infty)$, shown in Fig.~\ref{Data_Rate}. In the figure, we also provide the rate with optimal combiner ($\bw =  \frac{1}{\sqrt{N_\text{M}}} \boldsymbol{\alpha}_{r}(\phi_\text{R,M})$) and phase shifter values ($ \boldsymbol{\varphi} =  \frac{1}{\sqrt{N_\text{R}}} \boldsymbol{\alpha}_{t}(\theta_\text{R,M}) \odot \boldsymbol{\alpha}^*_{r}(\phi_\text{B,R})$), denoted by ``Optimal'' in the legend. Exhaustive search attains performance close to the optimal scheme. The random phase performs the worst among the four evaluated schemes, which indicates that the phase value also plays a crucial role in data transmission. 

\section{Conclusions}
We have considered a feasible yet efficient phase shifter design for the positioning and data transmission in the RIS-aided mmWave MIMO systems, where the LoS is obstructed. The RIS does not require any active elements and complicated computation, e.g., baseband processing units. We have evaluated the performance and compared it with random phase and exhaustive search schemes to show the superiority of the proposed scheme (achieving comparable performance to exhaustive search even in the low SNR regime with much less training overhead) and the importance of phase shifter design for both communication and positioning purposes. 
\balance
\bibliographystyle{IEEEbib}
\bibliography{IEEEabrv,Ref}

\end{document}